\documentclass[aps,prb,reprint,superscriptaddress,floatfix,titlepage]{revtex4-1}
\usepackage{graphicx}
\usepackage{amsmath}
\newcommand{\angstrom}{\mbox{\normalfont\AA}}

\begin{document}
\title{Tunable Electronic Structure in Gallium Chalcogenide van der Waals Compounds} 
\author{Brian Shevitski}
\thanks{Authors contributed equally.}

\affiliation{Department of Physics, University of California at Berkeley, Berkeley, California 94720,USA}
\affiliation{Molecular Foundry, Lawrence Berkeley National Laboratory, Berkeley, California 94720,USA}

\author{S{\o}ren~Ulstrup}
\email{ulstrup@phys.au.dk}
\thanks{Authors contributed equally.}

\affiliation{Advanced Light Source, E. O. Lawrence Berkeley National Laboratory, Berkeley,
California 94720, USA}

\affiliation{Department of Physics and Astronomy, Interdisciplinary Nanoscience Center (iNANO), Aarhus University, 8000 Aarhus C, Denmark}

\author{Roland J Koch}

\affiliation{Advanced Light Source, E. O. Lawrence Berkeley National Laboratory, Berkeley,
California 94720, USA}

\author{Hui Cai}
	\affiliation{School for Engineering of Matter, Transport and Energy, Arizona State University, Tempe, Arizona 85287, USA}
\author{Sefaattin Tongay}
	\affiliation{School for Engineering of Matter, Transport and Energy, Arizona State University, Tempe, Arizona 85287, USA} 	    
\author{Luca Moreschini}
\affiliation{Advanced Light Source, E. O. Lawrence Berkeley National Laboratory, Berkeley,
California 94720, USA}
\author{Chris Jozwiak}
\affiliation{Advanced Light Source, E. O. Lawrence Berkeley National Laboratory, Berkeley,
California 94720, USA}
\author{Aaron Bostwick}
\affiliation{Advanced Light Source, E. O. Lawrence Berkeley National Laboratory, Berkeley,
California 94720, USA}
\author{Alex Zettl}
    \affiliation{Department of Physics, University of California at Berkeley, Berkeley, California 94720,USA}
\author{Eli Rotenberg}
\affiliation{Advanced Light Source, E. O. Lawrence Berkeley National Laboratory, Berkeley,
California 94720, USA}

\author{Shaul Aloni}
\email{saloni@lbl.gov}

\affiliation{Molecular Foundry, Lawrence Berkeley National Laboratory, Berkeley, California 94720,USA}

\date{\today}
\begin{abstract}
Transition metal monochalcogenides comprise a class of two-dimensional materials with electronic band gaps that are highly sensitive to material thickness and chemical composition. Here, we explore the tunability of the electronic excitation spectrum in GaSe using angle-resolved photoemission spectroscopy. The electronic structure of the material is modified by $\textit{in-situ}$ potassium deposition as well as by forming GaS$_{x}$Se$_{1-x}$ alloy compounds. We find that potassium-dosed samples exhibit a substantial change of the dispersion around the valence band maximum (VBM). The observed band dispersion resembles that of a single tetralayer and is consistent with a transition from the direct gap character of the bulk to the indirect gap character expected for monolayer GaSe. Upon alloying with sulfur, we observe a phase transition from AB to $\text{AA}^{\prime}$ stacking. Alloying also results in a rigid energy shift of the VBM towards higher binding energies which correlates with a blue shift in the luminescence. The increase of the band gap upon sulfur alloying does not appear to change the dispersion or character of the VBM appreciably, implying that it is possible to engineer the gap of these materials while maintaining their salient electronic properties.
\end{abstract}

\maketitle

\section{Introduction}
The ability to isolate monolayers and study their physical properties has led to a resurgence of interest in layered van der Waals (VDW) materials as two-dimensional systems. Transition-metal chalcogenides belong to this class of materials and demonstrate unique thermoelectric, photonic and electronic properties\cite{wasala_recent_2017}. The composition, phase, and crystal structure of these materials can be tuned to display a wide range of electronic phases including metallic\cite{novoselov_electric_2004,feng_metallic_2011}, semiconducting\cite{mak_atomically_2010,splendiani_emerging_2010,watanabe_direct_2004}, superconducting\cite{guillamon_superconducting_2008,navarro_enhanced_2016,ugeda_layered_2015} and charge-density wave\cite{liu_low-frequency_2018,xi_two-dimensional_2015,ugeda_layered_2015}, inspiring utilization in a wide range of applications including electronic devices, sensors, and quantum devices\cite{choi_recent_2017}.
\par
In their bulk form, monochalcogenides of Ga and In,  such as GaSe, InSe, and GaS are van der Waals compounds.  Gallium chalcogenides (GCs) are direct bandgap semiconductors that can be mechanically exfoliated from a bulk parent crystal. Their properties are highly sensitive to material thickness, chemical composition, and crystalline structure\cite{rybkovskiy_transition_2014,cai_band_2016,robertson_electronic_2001}. Bulk GaSe and GaS have direct band gaps of 2.10 eV and 3.05 eV, respectively, and transition to indirect gaps when the number of layers is reduced\cite{lei_synthesis_2013,aziza_tunable_2017}. 
Monolayer field effect devices maintain mobilities on the order of of 0.1 cm$^2$V$^{-1}$s$^{-1}$, on/off ratios of $10^4$ - $10^5$ as well as quantum efficiencies that exceed those of graphene by several orders of magnitude when implemented in a device architecture\cite{hu_synthesis_2012,zhou_epitaxy_2014,late_gas_2012}. 
\par
Alloys with mixed chalcogen content, GaS$_{x}$Se$_{1-x}$, allow for control of the bandgap, allowing engineering of the optical absorption and photoluminescence (PL)\cite{serizawa_polytypes_1980,jung_red--ultraviolet_2015}. 
Interlayer coupling, defects, doping, and interaction with an underlying substrate may further modify the electronic structure of the GCs, as observed in exfoliated flakes of GaSe and GaTe on SiO2 and graphene\cite{chen_intrinsic_2015,del-pozo_photoluminescence_2015,kim_tunable_2016} and epitaxially grown GaSe flakes on graphene\cite{aziza_tunable_2017,aziza_van_2016,li_van_2015}.
\par
In this communication, we illustrate how the crystalline and electronic properties of GC materials are affected by changes in chemical composition, doping, and reduced dimensionality using high resolution scanning transmission electron microscopy (HRSTEM), angle-resolved photoemission spectroscopy (ARPES), and PL. 
\par
\begin{figure*}[ht]
	\centering
	\includegraphics[width=0.98\textwidth]{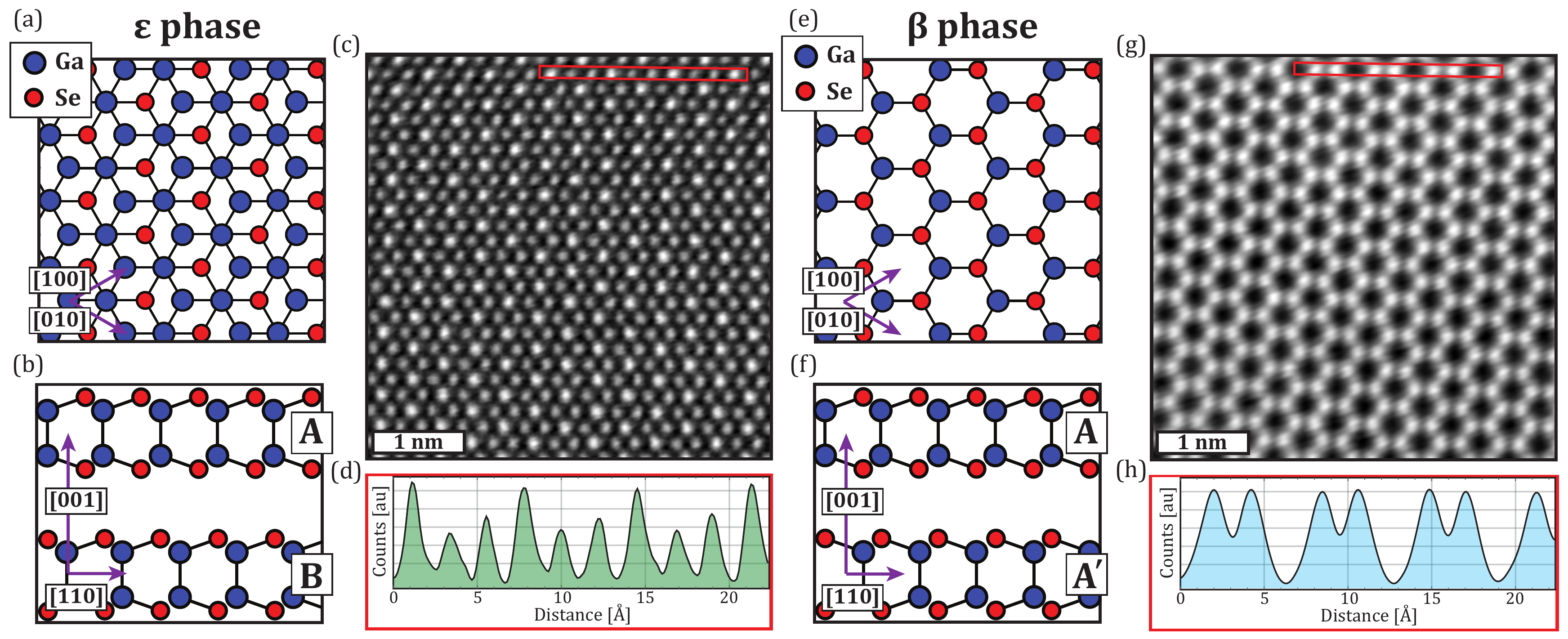}
	\caption{(color online) Real space structure and HRSTEM images of the two most common GaS$_{x}$Se$_{1-x}$ polytypes. Panels (a) and (b) show the crystal structure of the $\varepsilon$ phase in the planes defined by the axes in the labels. The corresponding HRSTEM image in (c) shows a trigonal lattice, resulting from the projection of Bernal (AB) stacked honeycomb layers. The intensity cut in (d) was obtained from the red rectangle in (c) and shows bright atomic columns with approximately twice the intensity of dim atomic columns. Analogous images in (e)-(f) show the geometry of the $\beta$ phase. The HRSTEM image in (g) shows a honeycomb mesh, with no intensity in the interstitial columns between hexagonal cells.}
	\label{fig:structure}
\end{figure*}
HRSTEM reveals that GaSe tetralayers stack in the AB sequence ($\varepsilon$ phase) while sulfur alloys and GaS stack almost exclusively in the AA$^{\prime}$ sequence ($\beta$ phase). The complete $E\left(k_x,k_y,k_z\right)$ electronic structure of AB-stacked GaSe is determined using ARPES, allowing us to directly extract the band dispersion along high symmetry directions and determine valence band (VB) extrema and effective masses. We illustrate how the full energy-momentum-resolved quasiparticle band structure of the GCs is modulated when the chemical composition and/or interaction between layers is changed. We explore the influence of potassium deposition on the band structure, finding it only causes a slight energy shift of the VBs and does not lead to degenerate doping which would enable ARPES visualization of the conduction band. However, potassium dosing does lead to a significant change of the dispersion around the top of the VB, implying a strong modification at the surface GaSe tetralayer compared to the bulk. Finally, we explore the effect of sulfur alloying on the band structure. We find that increasing sulfur content produces a rigid VB shift and results in a corresponding blue shift in the PL, suggesting that alloying is a viable method for tuning GC optical properties.
\section{Experimental}
We study pure GaSe and GaS as well as three alloy compositions spanning the entire compositional range of GaS$_{x}$Se$_{1-x}$ alloys. Crystals are grown using the modified Bridgman-Stockbarger method\cite{ni_growth_2013}. Briefly, stoichiometric amounts of Ga, S, and Se powder are weighed and mixed in sealed quartz ampoules to achieve the desired alloying ratio. The Ampoules are then heated in a zone furnace at 970 $^{\circ}$C for two weeks to grow single crystals 2-8 mm in diameter. 
\par
The HRSTEM samples are prepared by a two-step process. First, the crystals are cleaved using blue wafer dicing tape and transferred to Si substrates coated with 100 nm thick SiO$_2$. Next, a Cu TEM grid with carbon film is adhered to flakes on the Si/SiO$_2$ wafer using a drop of isopropanol. Once the drop has dried, the SiO$_2$ is etched away using 1M NaOH. Lastly the grid is washed in deionized water and dried prior to imaging. HRSTEM is carried out using a double $C_s$ corrected FEI Titan 80-300 operating at an accelerating voltage of 80 kV at the Molecular Foundry. 
\par
For ARPES measurements, large crystals of GCs are glued to a Cu sample holder using epoxy and cleaved $\textit{in-situ}$ in the ultra-high vacuum (UHV) chamber with a base pressure better than 5$\times 10^{-11}$ mbar. The samples are cooled to 85 K prior to cleaving and kept at this temperature during ARPES measurements. The cleaved single-crystal domain sizes are on the order of 100 $\mu$m as defined by spatially scanning the sample with the synchrotron beam. The ARPES data are collected at the Microscopic and Electronic Structure Observatory (MAESTRO) at the Advanced Light Source (ALS) using the microARPES end-station equipped with a hemispherical Scienta R4000 analyzer. The beamline slit settings are adjusted so that the size of the beam is on the order of 20 $\mu$m. VB and Ga 3d core level spectra are obtained primarily using a photon energy of 94 eV. S 2p and Se 3p core level data are collected using a photon energy of 300 eV. Photon energy scans of GaSe are acquired for photon energies between 21 eV and 140 eV. In order to relate the photoelectron kinetic energy, E$_{kin}$, to the out-of-plane momentum, $k_z$, we use the free-electron final state assumption where $k_z^2 = \left(2m/\hbar^2\right) \left(\text{E}_{kin} + \text{V}_0\right)$, where V$_0$ is the inner potential\cite{hufner_2003}. We find that $\text{V}_{0} = 10.2$  eV provides the best description of the data, given the out-of-plane lattice constant c = 15.96 $\angstrom$ and the Brillouin zone (BZ) periodicity of $2\pi/\text{c}$ for the measured polytypes of GaSe. 
\par
\begin{figure*}[th]
	\centering
	\includegraphics[width=0.9\textwidth]{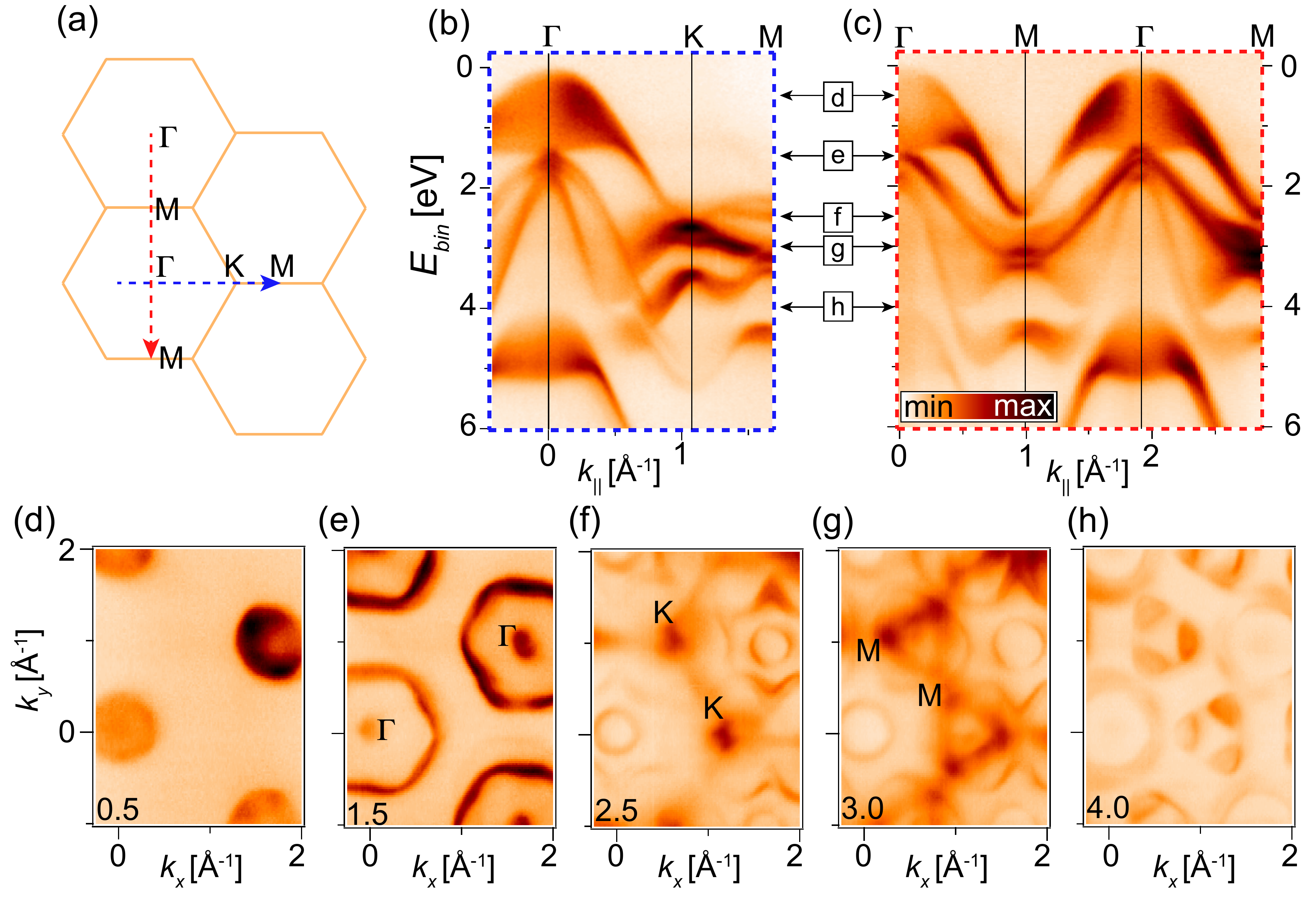}
	\caption{(color online) ARPES measurements at a photon energy of 94 eV of the electronic structure of pristine GaSe around the $\Gamma$ point. (a) Sketch of the BZs and their orientation in the experiment with $\Gamma$, $K$ and $M$ symmetry points labeled.(b) Measured dispersion in the $\Gamma-K-M$ direction obtained along the blue dashed line shown in (a). (c) Measured dispersion in the $\Gamma-M-\Gamma$ direction obtained along the red dashed line shown in (a). The double headed arrows in (b)-(c) indicate the planes of constant energy from which the cuts in (d)-(h) are taken. The energy of each cut is also listed in bottom of each panel in units of electron volts (eV). High symmetry points $\Gamma$, $K$ and $M$ have been labeled in (e), (f) and (g), respectively.}
	\label{fig:inplane_dispersion}
\end{figure*}
The Fermi energy is determined on the clean Cu sample holder in contact with the crystals for all photon energies. The presence of surface photovoltage-induced energy shifts in the samples is checked by varying the photon flux. No such changes were observed, except on GaS, which excluded ARPES measurements on this material.
\par
Potassium dosing experiments on GaSe are carried out $\it{in-situ}$ using SAES getters mounted in the analysis chamber while the sample is kept at 85 K. Each dose takes 50 seconds, and between doses, an $(E,k)$ spectrum of the VB and an energy spectrum over the Ga 3d/K 3p binding energy region are taken. The collection of these spectra takes 70 seconds and 12 doses were applied for the results presented here, amounting to a total time of 24 minutes for collecting the dataset.  
\par
Judging from the change in photoemission intensity from the VB upon further dosing, we estimate a coverage rate of a complete monolayer after 12 doses. We observe that prolonged exposure to the synchrotron beam for the fully K-dosed sample leads to broadening and significant deterioration of the quality of the spectral features, preventing detailed photon energy or angle scans for the K-dosed samples.
\par
The total energy and momentum resolution in the ARPES data are better than 20 meV and 0.01 $\angstrom^{-1}$, respectively.
\section{Results and Discussion}
\subsection{HRSTEM Measurements of Gallium Chalcogenides}
The GaS$_{x}$Se$_{1-x}$ system forms a continuous set of alloys for the entire compositional range ($0\leq x \leq 1$) of the stoichiometry\cite{whitehouse_structural_1978,terhell_polytype_1982,arancia_electron_1976}. Like other layered VdW materials, the intralayer bonding is strong and mostly covalent while the interlayer bonding is weak and of VdW character. The unit cell consists of two tetrahedral pyramids stacked tip to tip with chalcogen atoms at the base and a Ga atom at the apex, forming a four-atom-thick tetralayer. 
\par
The layer stacking exhibits several different polytypes, depending on composition and growth conditions. In the $\varepsilon$ phase (Fig. \ref{fig:structure}(a-b)), successive tetralayers are stacked in the sequence AB, resulting in Ga atoms from the A layer being vertically aligned with chalcogen atoms in the the B layer, and chalcogen atoms in the A layer being vertically aligned with the interstitial voids in the centers of the hexagons from the B layer, similar to Bernal stacking in graphite. The inversion symmetric $\beta$ phase (Fig. \ref{fig:structure}(e-f)) is obtained by stacking successive tetralayers in the sequence AA$^{\prime}$, resulting in Ga(chalcogen) atoms from the A layer eclipsing chalcogen(Ga) atoms from the A$^{\prime}$ layer. The system can also form the rhombohedral $\gamma$ phase (ABC stacking), but this stacking sequence is not observed experimentally in our samples.
\par
Previous studies have explored the relationship between sulfur content and stacking phase\cite{whitehouse_structural_1978,terhell_polytype_1982,arancia_electron_1976}. GaS$_{x}$Se$_{1-x}$ alloys with low sulfur content (x $<$ 0.15) stack in the $\varepsilon$ phase while alloys with higher sulfur content (x $>$ 0.35) display the $\beta$ stacking phase. Our experimental results (described below) also show this trend. We find that the $\varepsilon$ phase occurs for x $\leq$ 0.3 and the $\beta$ phase for x $>$ 0.3,  The $\beta$ and $\varepsilon$ phases are thought to co-exist for intermediate sulfur content (0.15 $<$ x $<$ 0.35), phase segregated throughout the crystal, but our samples do not show experimental evidence for this theory.
\par
Contrast in HRSTEM indicates differences in atomic number, density, and thickness throughout a specimen. Due to the specific stacking sequences exhibited by the polytypes of GaS$_{x}$Se$_{1-x}$ alloys and the contrast mechanisms of STEM imaging, our imaging experiments (Fig. \ref{fig:structure}(c) and \ref{fig:structure}(g)) unambiguously reveal the stacking phases of these materials.
\par
Figure \ref{fig:structure}(c) shows a typical HRSTEM image for an alloy with high sulfur content (x $\sim 0.75$), appearing as a trigonal lattice with alternating bright and dim atomic columns. Figure \ref{fig:structure}(d) shows a line plot of the intensity extracted from the region outlined by the red rectangle in Fig. \ref{fig:structure}(c). The heights of the peaks in the line trace indicate that there are approximately twice as many atoms in bright columns compared to the dim columns, consistent with the stacking geometry of the $\varepsilon$ phase, shown in Fig. \ref{fig:structure}(b). Figure \ref{fig:structure}(g) shows an analogous image for pure GaSe which appears as a honeycomb mesh of atomic columns with no appreciable intensity in the interstitial voids between hexagons. The line profile (Fig. \ref{fig:structure}(h)) shows that the atomic columns have uniform intensity, consistent with AA$^{\prime}$ stacking exhibited by the $\beta$ phase. 
\subsection{ARPES Electronic Structure Measurements of GaSe}
An overview of the electronic band structure of pristine GaSe is presented in Fig. \ref{fig:inplane_dispersion}. The data are acquired at a photon energy of 94~eV, which probes around the $\Gamma$ point, as described by the BZ sketches in Fig. \ref{fig:inplane_dispersion}(a). Figures \ref{fig:inplane_dispersion}(b)-(c) show cuts along the $\Gamma -  K - M$ and $\Gamma -  M - \Gamma$ high symmetry directions as signified by the dashed lines in Fig. \ref{fig:inplane_dispersion}(a). The evolution of the dispersion with binding energy is presented via the constant binding energy cuts in Fig. \ref{fig:inplane_dispersion}(d)-(h), which reveal the trigonal symmetry of the dispersion as well as strong intensity variations between the bands in the BZs. The valence band maximum (VBM) is situated at $\Gamma$ and is characterized by a parabolic-shaped lobe. The location of the VBM at $\Gamma$ is consistent with theoretical predictions that GaSe is a direct-gap semiconductor with the conduction band minimum (CBM) also located at $\Gamma$ \cite{olguin_ab-initio_2013}. Additional sub bands dispersing towards $M$ and $K$ appear above binding energies of 1.5~eV. The material remains rather featureless away from $\Gamma$ before the onset of this higher binding energy range, implying most of the optical properties of GaSe will derive from the central parabolic lobe. The ARPES data discussed here are fully consistent with the electronic structure data presented in Ref. \citenum{plucinski_electronic_2003}.
\par
\begin{figure}[tb]
	\centering
	\includegraphics[width=0.49\textwidth]{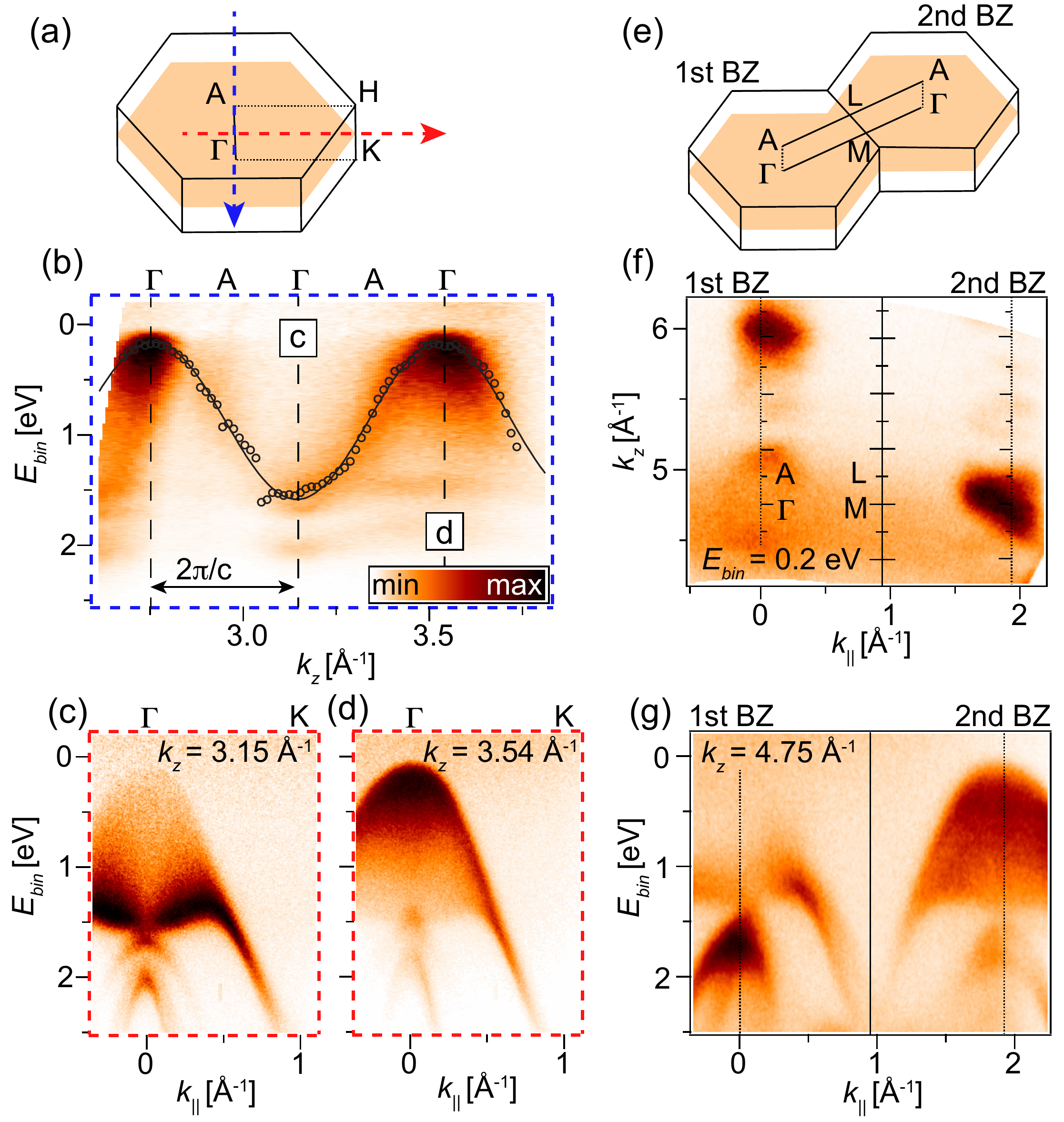}
	\caption{(color online) ARPES measurements of the $k_z$-dispersion of GaSe. (a) Sketch of bulk BZ highlighting the $\Gamma - A - H - K$ plane. (b) Dispersion along the $\Gamma - A$ direction, as marked by the blue dashed line in (a). Fits of the position of the VBM intensity are shown via open circles. The solid curve following the fitted dispersion is a guide for the eye. The BZ size of $2\pi/\text{c}$ in the $z$-direction is marked by a double-headed arrow and two vertical dashed lines. (c)-(d) Dispersion extracted at the given $k_z$ values, corresponding to the $\Gamma - K$ direction (red dashed line). (e)  Sketch of 1st and 2nd BZs with an outline of the $\Gamma - A - L - M$ plane. (f) Constant energy contour of the plane sketched in (e) extracted at a binding energy of 0.2 eV with high symmetry points annotated by tick marks. (g) Dispersion along the $\Gamma - M - \Gamma$ direction obtained at a $k_z$ value of 4.75~\AA$^{-1}$.}
	\label{fig:outofplane_dispersion}
\end{figure}
We investigate the $k_z$-dispersion in the three dimensional BZ of GaSe shown in Fig. \ref{fig:outofplane_dispersion}(a). By measuring the photon energy dependence of the dispersion at normal emission we are able to trace the band structure along the $\Gamma - A$ direction, as seen in Fig. \ref{fig:outofplane_dispersion}(b). A strong $k_z$-dispersion is visible in the center of the parabolic lobe, which is further highlighted via the fits (open circles) and guide line in Fig. \ref{fig:outofplane_dispersion}(b), whereas the bands at binding energies higher than 1.5~eV have a much more flat $k_z$-dispersion. The fitted dispersion reveals an apparent BZ doubling effect; however, this occurs due to a suppression of the photoemission intensity in adjacent BZs\cite{plucinski_electronic_2003}, similar to observations in photoemission experiments on graphite\cite{shirley_brillouin_1995,mucha_characterization_2008}. The band structure along the $\Gamma - K$ line is shown for two neighboring BZs in the $k_z$ direction in Fig. \ref{fig:outofplane_dispersion}(c-d). The main intensity moves from the bottom corners of the parabolic lobe in Fig. \ref{fig:outofplane_dispersion}(c) to the VBM in Fig. \ref{fig:outofplane_dispersion}(d). Similar intensity variations are observed in neighboring in-plane BZs, as sketched in Fig. \ref{fig:outofplane_dispersion}(e). The ($k_{||}$, $k_z$)-contour at a binding energy of 0.2~eV shown in Fig. \ref{fig:outofplane_dispersion}(f) shows a cut around the VBM for the $\Gamma - A - L - M$ plane (see Fig. \ref{fig:outofplane_dispersion}(e)). The intensity maxima are seen to shift between the 1st and 2nd BZs and the absolute intensity levels also vary strongly within the same BZ. A cut along the $\Gamma - M$ line at $k_z$ = 4.75 \AA$^{-1}$ in Fig. \ref{fig:outofplane_dispersion}(g) shows the upper VB parabolic lobe in two neighboring zones. In the first BZ, the intensity is concentrated in the bottom corner while in the 2nd BZ the intensity is concentrated around the VBM. 
\par
\begin{figure}[tb]
	\centering
	\includegraphics[width=0.49\textwidth]{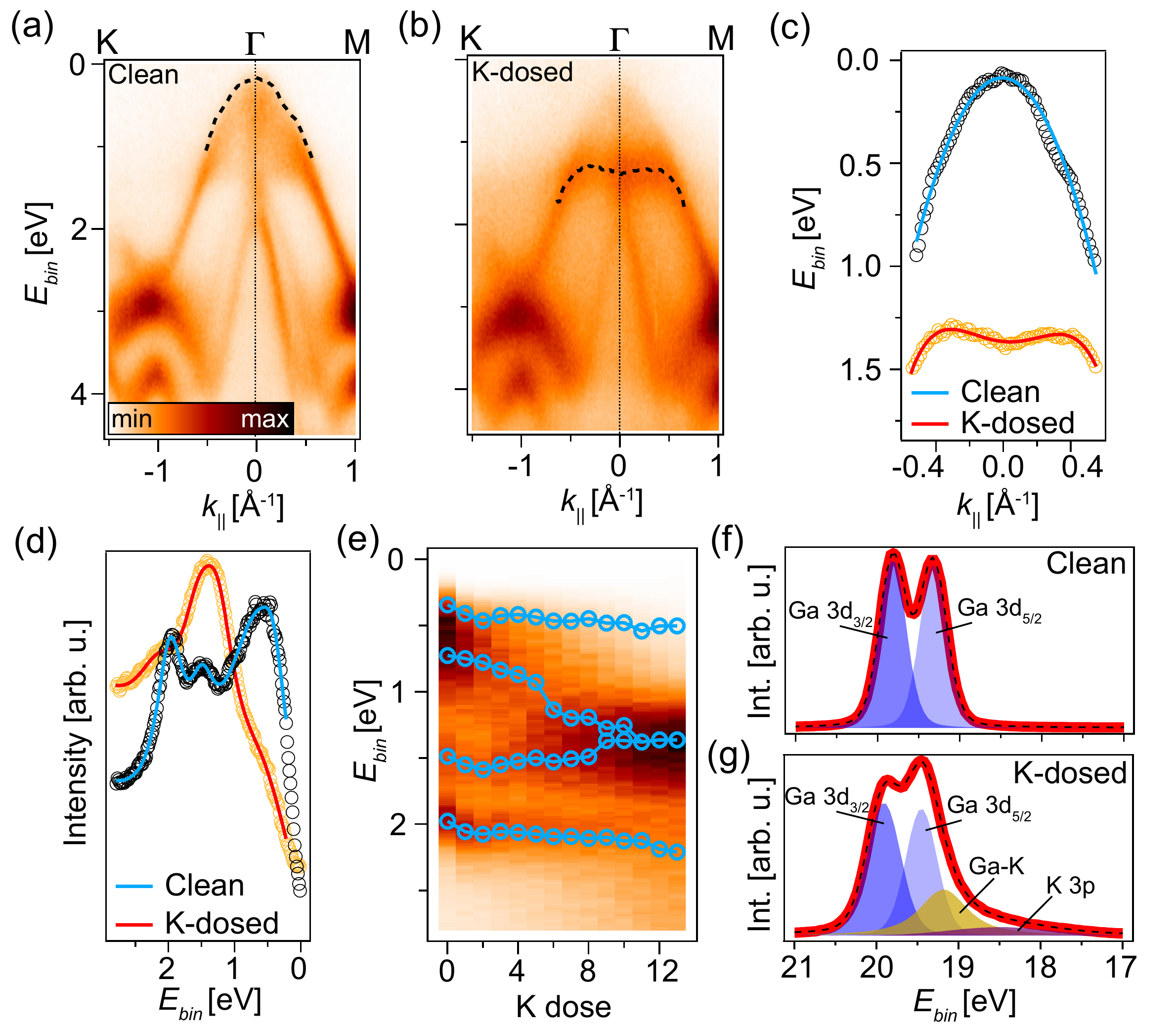}
	\caption{(color online) Effect of K dosing on the GaSe electronic structure. (a)-(b) ARPES Measurements of the dispersion in the $\Gamma - K - M$ direction (a) before and (b) after K dosing. The dashed curves present the fitted location of the VBM peak intensity. (c) Location of the VBM peak intensity (open circles) fitted with a 2nd order and a 4th order polynomial for the clean and K dosed samples, respectively. (d) EDCs (open circles) and results of Voigt function fits (curves) in the clean and K dosed cases. The EDCs were extracted at $\Gamma$ along the lines shown in (a)-(b). The position of the top-most peak is plotted via dashed lines in (a)-(b) and open circles in (c). (e) Evolution of the photoemission intensity at $\Gamma$ as a function of K dosing steps. The open circles and guidelines mark the fitted peak positions obtained by performing the EDC analysis shown in (d) in each K dosing step. (f)-(g) Ga 3d core level measurements (f) before and (g) after K dosing. The data (thick red curves) have been fitted (dashed curves) with Doniach-Sunjic line shapes (filled peaks) for each of the observed components. }
	\label{fig:dosing}
\end{figure}
This behavior can be reconciled with the orbital character of the bands, previously studied with detailed calculations\cite{rybkovskiy_transition_2014}. Since the states near the VBM are mainly of Se $p_z$ character, the interlayer interactions will cause intensity modulations reflecting the coupling of the GaSe tetralayers, leading to a dispersion resembling a single tetralayer for the $k_z$ value used for the slice shown in Fig. \ref{fig:outofplane_dispersion}(c), while the $k_z$ value for Fig. \ref{fig:outofplane_dispersion}(d) shows a dispersion with a more bulk-like character\cite{rybkovskiy_transition_2014}. The subbands just below the VBM exhibit less dramatic behavior due to the $p_{x(y)}$ character of these states which are less sensitive to the adjacent GaSe tetralayers\cite{camara_electronic_2002,nagel_tight_1979}.
\subsection{Potassium Deposition of GaSe}
The effect of K dosing of bulk GaSe is explored in Fig. \ref{fig:dosing}. Alkali metal deposition offers a possibility to chemically donate electrons near the surface of GaSe and thereby dope the material. The ARPES measurements along the $\Gamma - K - M$ direction in Fig. \ref{fig:dosing}(a-b) before and after complete K dosing show that a significant band structure change occurs, rather than strong electron doping. A maximum energy shift of 0.1 eV of the VB towards higher binding energies is observed, which may originate from surface band bending due to the added charge or a slight doping of the material. After further dosing, we observe that the spectral weight shifts from the top (see Fig. \ref{fig:dosing}(a)) to the bottom (see Fig. \ref{fig:dosing}(b)) of the VB lobe. In Fig. \ref{fig:dosing}(c), the VBM dispersion (open circles), extracted via fits to energy distribution curves (EDCs), has been fitted with 2nd and 4th order polynomial functions in the clean and K doped cases, respectively. In the pristine GaSe, a global VBM is observed at $\Gamma$ and at a binding energy of (0.09 $\pm$ 0.03) eV. The effective mass around the top of the parabola is $m^*$ = (1.1 $\pm$ 0.2)$m_0$, where $m_0$ is the free electron mass. In K dosed GaSe, the VBM splits off into two maxima located at $\pm$0.3 \AA$^{-1}$ with a binding energy of (1.30 $\pm$ 0.04) eV and an effective mass of $m^*$ = (1.7 $\pm$ 0.2)$m_0$. 
\par
The EDCs extracted at normal emission in Fig. \ref{fig:dosing}(d) trace the peak positions at each K dosing step. Voigt line shapes have been fitted to each of the peaks, which have been plotted together with the photoemission intensity at normal emission at each K dose in Fig. \ref{fig:dosing}(e), revealing a gradual shift of the ARPES intensity from the pristine bulk dispersion seen in Fig. \ref{fig:dosing}(a) to the modified dispersion shown in Fig. \ref{fig:dosing}(b). 
The modified dispersion strongly resembles the ``bow-shape'' or``inverted sombrero'' dispersion of single-layer GaSe, where the inversion of the VB is characterized by the energy difference between the VBM and the local energy minimum at $\Gamma$\cite{rybkovskiy_transition_2014,jung_red--ultraviolet_2015,rybkovskiy_size_2011,ben_aziza_valence_2018,chen_large-grain_2018}. Here, we find that the band inversion is (48 $\pm$ 12) meV, which compares well with a value of 80 meV obtained from recent DFT calculations\cite{ben_aziza_valence_2018}. The value is substantially smaller than 120 meV found for single-layer GaSe grown on graphene on silicon carbide\cite{chen_large-grain_2018} and than the value of (150 +/- 10) meV that we determine for the similar dispersion at $k_z = 3.15$  \AA$^{-1}$ in Fig. \ref{fig:outofplane_dispersion}(c). We expect that interlayer interactions between the Se $p_z$ orbitals play a significant role for these dispersion changes around the VBM\cite{rybkovskiy_transition_2014} and that the gradual change observed in Fig. \ref{fig:dosing}(e) signals a modification of the topmost GaSe layer, leading to a weaker coupling to the underlying bulk. This may be facilitated by K intercalation in the van der Waals gap between the layers, as observed in metallic\cite{rossnagel_continuous_2005}, and semiconducting\cite{eknapakul_electronic_2014} transition metal dichalcogenides, or by a chemical interaction on the surface of the crystal, which significantly changes the coupling between individual layers as seen in graphene multilayers\cite{ohta_controlling_2006,ohta_interlayer_2007}. In these situations one may also expect a change of the interlayer separation of tetralayers as well as a shift of inner potential, which also contribute to the spectral changes we observe. The possibility of chemical changes near the surface is further supported by the observation of an additional core level component in the binding energy range between the Ga 3d and K 3p core levels for the K-dosed sample, which is demonstrated by the comparison between the clean GaSe core level spectrum in Fig. \ref{fig:dosing}(f) and the spectrum for the fully K-dosed sample in Fig. \ref{fig:dosing}(g). 
\begin{figure}[b]
	\centering
	\includegraphics[width=0.35\textwidth]{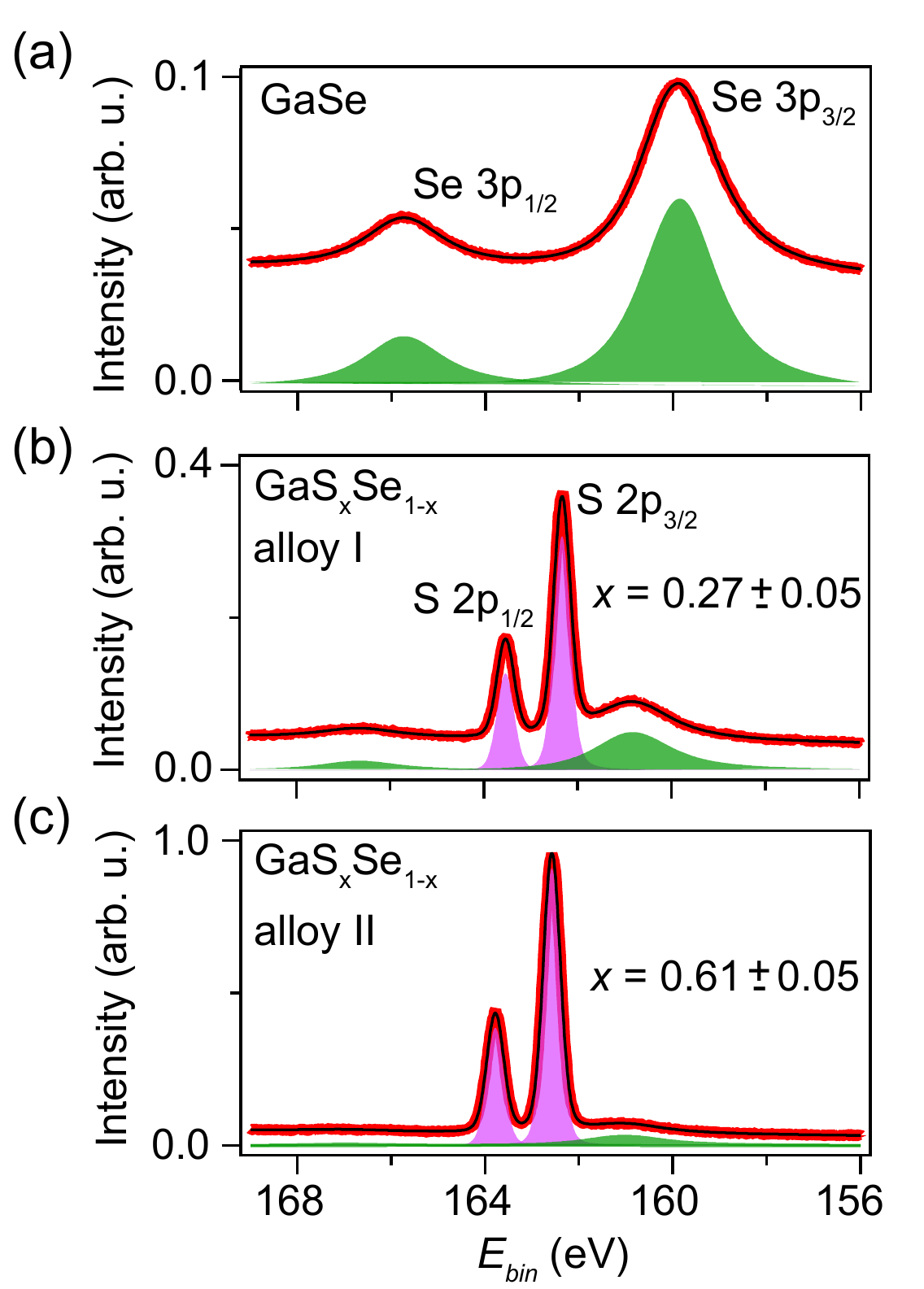}
	\caption{(color online) Composition of GaS$_x$Se$_{1-x}$ alloys from core level data. (a)-(c) Core level measurements (red thick curves) in the Se 3p and S 2p binding energy region for (a) GaSe and (b)-(c) GaS$_x$Se$_{1-x}$ alloys. Fit results (black curves) to Doniach-Sunjic line shapes (filled peaks) are provided in each panel. The composition $x$ of the alloys determined via the fitted peak components is stated in (b)-(c). Note the intensity scale varies between the panels.}
	\label{fig:alloys_xps}
\end{figure}
\subsection{ARPES and PL Measurements of GaS$_x$Se$_{1-x}$}
GaS$_{x}$Se$_{1-x}$ alloys offer a promising route to tune the optical gap and luminescent properties of III-VI semiconductors. However, it is unknown how robust the electronic structure remains between different alloys. Here we compare pristine GaSe with two different GaS$_{x}$Se$_{1-x}$ alloys (alloys I and II). The stoichiometry $x$ is checked via core level measurements of the Se 3p and S 2p binding energy regions, shown in Fig. \ref{fig:alloys_xps}. By comparing the spectral weight of the Se 3p core levels, we are able to estimate that the composition of alloy I is x = 0.27 $\pm$ 0.05 and the composition of alloy II is x = 0.61 $\pm$ 0.05. These values are consistent when we perform the same analysis on the Se 3d core levels (not shown). The core level data is in agreement with the stoichiometry of precursor added to the growth ampoule. 
\par
\begin{figure}[b]
	\centering
	\includegraphics[width=0.49\textwidth]{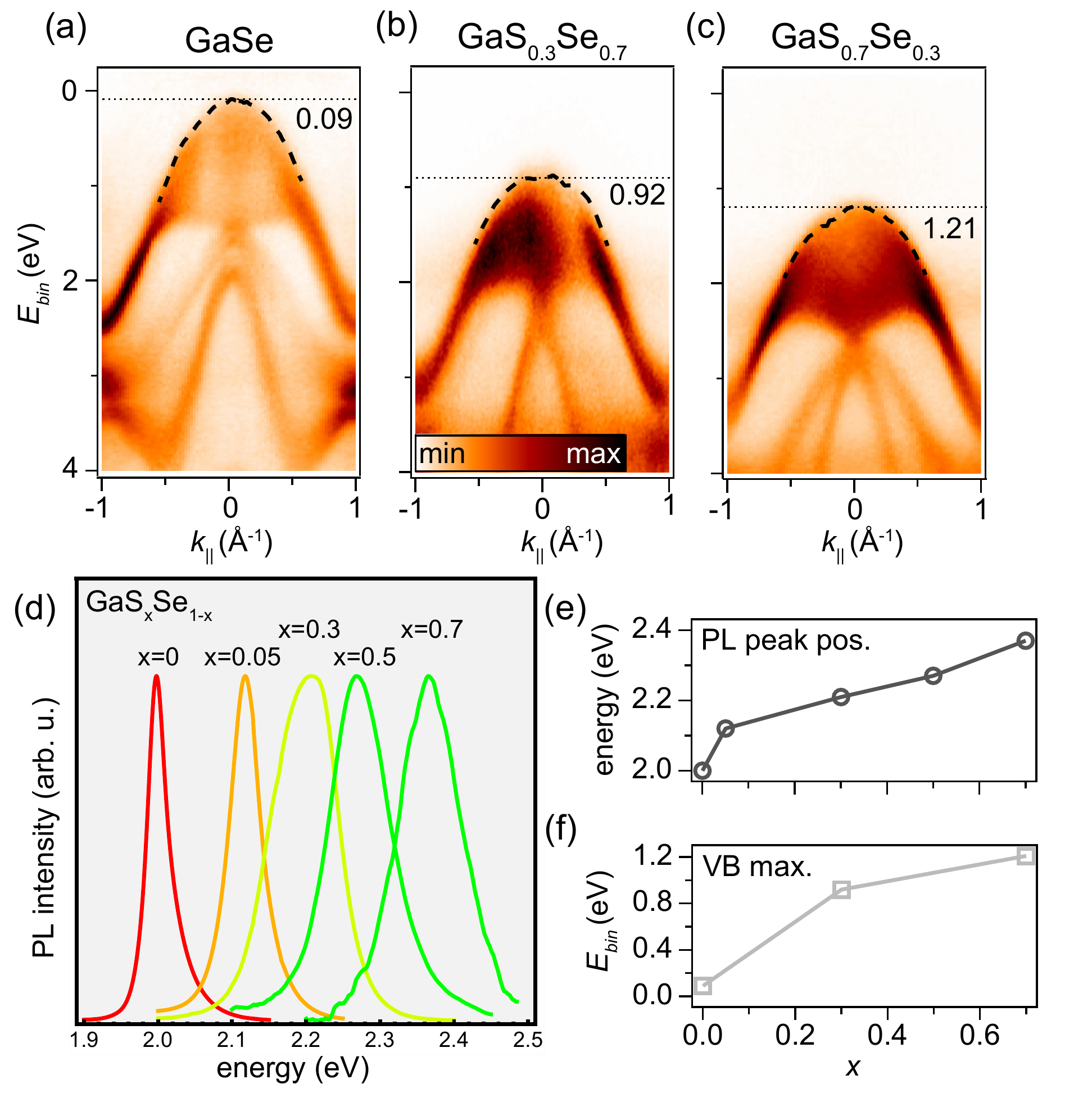}
	\caption{(color online) VBM and PL peaks positions for GaS$_x$Se$_{1-x}$ alloys. (a)-(c) ARPES measurements of the upper VB region for (a) GaSe, (b) GaS$_{0.27}$Se$_{0.73}$ and (c) GaS$_{0.61}$Se$_{0.39}$. The dotted line presents the fitted position of the VBM and the dashed curves exhibit the fitted dispersion around the VBM in each case. The corresponding VBM binding energy positions are given in units of eV with an error bar of $\pm 0.03$~eV. (d) PL intensity for the same samples as in (a)-(c) and Fig. \ref{fig:alloys_xps}. (e)-(f) Positions of the (e) PL peaks and (f) VBM as a function of composition.}
	\label{fig:alloys_arpes}
\end{figure}
Figure \ref{fig:alloys_arpes}(a)-(c) present ARPES measurements of the VBs for the three systems. The overall band dispersion is found to be very similar with a constant effective mass of $m^*$ = (1.1 $\pm$ 0.2)$m_0$. The binding energy position of the VBM changes significantly between the three systems, indicating a change of the electronic band gap. It is also possible that Fermi level pinning plays a strong role for the band positions, as the defect concentration is likely different in the alloys. PL measurements, shown in Fig. \ref{fig:alloys_arpes}(d), reveal a change of the optical gap. The PL peak positions plotted in Fig. \ref{fig:alloys_arpes}(e) show that the optical gap increases as more S is added in the alloy. The same trend is seen for the VBM binding energy position in Fig. \ref{fig:alloys_arpes}(f), which fits with a simultaneous increase of the quasiparticle band gap. The trend determined here is fully consistent with optical absorption measurements of GaS$_{x}$Se$_{1-x}$\cite{jung_red--ultraviolet_2015}. Note that ARPES measurements of pure GaS were attempted but the sample was found to charge too severely, which is likely due to the larger band gap of the material compared to the other GaS$_{x}$Se$_{1-x}$ compounds investigated here. 
\par
The $\varepsilon$ and $\beta$ stacking phases observed in HRSTEM are not expected to give rise to striking differences in the electronic structure\cite{camara_electronic_2002,nagel_tight_1979,rybkovskiy_size_2011}. However, detailed inspection of the measurements in Fig.. \ref{fig:alloys_arpes}(a)-(c) does show some variation in the intensity and structure of the dispersive bands in the higher binding energy region above the central VB lobe. While different stacking phases may affect the photoemission intensity, the change in chemical environment caused by the mixed chalcogen content directly influences the in-plane p$_{x(y)}$ orbitals that the higher binding energy bands are composed of, thereby leading to the possibility of a modified dispersion.

\section{Conclusion}
We have measured the bulk electronic structure of GaSe using ARPES and thereby identified the bulk band dispersion, the structure of the VBM, and the complex photoemission intensity variations associated with the tetralayer unit cell. Potassium deposition of GaSe leads to a dramatic change of the VBM dispersion from a parabolic shape characteristic for the bulk to an inverted bow-like shape that is usually associated with single-layer samples. Our data are consistent with an increased band gap and a transition from a direct to an indirect band gap semiconductor, which is caused by a strong modification of the top-most GaSe tetralayer due to the interaction with the deposited K atoms. We have investigated the effect of sulfur alloying on the crystalline and electronic structure and used HRSTEM to observe a transition from the $\varepsilon$ phase to the $\beta$ phase for sulfur content above $\sim$30 percent. We have also shown that alloying causes a rigid shift of the VBM binding energy position without a significant change in the actual band dispersion around the VBM. This rigid shift is consistent with an increase in the optical band gap measured by PL.

\begin{acknowledgments} 
This work was supported primarily by the U.S. Department of Energy, Office of Science, Office of Basic Energy Sciences, Materials Sciences and Engineering Division, under Contract No. DE-AC02-05CH11231, within the van der Waals Heterostructures Program (KCWF16), which provided for HRSTEM and ARPES support.  This work was also supported in part by the National Science Foundation under Grant No. DMR-1807233, which provided for PL measurements and preliminary exploration of sample growth parameters.  S.U. acknowledges financial support from VILLUM FONDEN under the Young Investigator Program (Grant No. 15375). R.J.K. was supported by a fellowship within the Postdoc-Program of the German Academic Exchange Service (DAAD). The Advanced Light Source is supported by the Director, Office of Science, Office of Basic Energy Sciences, of the U.S. Department of Energy under Contract No. DE-AC02-05CH11231. Work at the Molecular Foundry was supported by the Office of Science, Office of Basic Energy Sciences, of the US Department of Energy under Contract No. DE-AC02-05CH11231. B.S. acknowledges support from the NSF LSAMP BD graduate fellowship.
\end{acknowledgments}

\bibliography{GaSeS_ARPES.bib}

\end{document}